\documentclass[aps,prl,preprint,tightenlines,superscriptaddress,showpacs,byrevtex]{revtex4}

\usepackage{graphicx} % Include figure files
\usepackage{dcolumn}  % Align table columns on decimal point
\usepackage{color}

\graphicspath{{ps}}

\newcommand{\Kpipi}{$\tau^{-}\rightarrow K^{-} \pi^{+}\pi^{-}\nu_{\tau}$}
\newcommand{\pipipi}{$\tau^-\rightarrow \pi^-\pi^+\pi^-\nu_{\tau}$}
\newcommand{\KKpi}{$\tau^-\rightarrow K^-K^+\pi^-\nu_{\tau}$}
\newcommand{\KKK}{$\tau^-\rightarrow K^-K^+K^-\nu_{\tau}$}
\newcommand{\Kspi}{$\tau^-\rightarrow K_S^0\pi^-\nu_{\tau}$}
\newcommand{\enunu}{$\tau\rightarrow e\overline{\nu}\nu$}
\newcommand{\mununu}{$\tau\rightarrow \mu\overline{\nu}\nu$}
\newcommand{\pipipipizero}{$\tau^-\rightarrow \pi^-\pi^+\pi^-\pi^0\nu_{\tau}$}
\newcommand{\eeqq}{$e^+e^-\rightarrow q\bar{q}$}

\newcommand{\MKpipi}{$M(K\pi\pi)$}
\newcommand{\Mpipipi}{$M(\pi\pi\pi)$}
\newcommand{\MKKpi}{$M(KK\pi)$}
\newcommand{\MKKK}{$M(KKK)$}

\newcommand{\Ks}{\textcolor{black}{$K_S^0$}}
\newcommand{\pizero}{$\pi^0$}
\newcommand{\fbi}{$\rm {fb^{-1}}$}
\newcommand{\plusminus}{$\pm$}

\newcommand{\GeVc}{GeV/$c$}

\newcommand{\GeVcc}{GeV/$c^2$}
\newcommand{\MeVcc}{MeV/$c^2$}
\newcommand{\emuevent}{\{$\,e,\mu\,$\}}
\newcommand{\eetautau}{$e^+e^-\rightarrow \tau^+\tau^-$}
\newcommand{\degree}{$^{\circ}$}

\newcommand{\update}[1]{\textcolor{black}{#1}}
\newcommand{\updatea}[1]{\textcolor{black}{#1}}
\newcommand{\updateb}[1]{\textcolor{black}{#1}}

\begin{document}

\preprint{\vbox{ \hbox{   }
						\hbox{Belle--CONF--{\it 0809}}
}}

\title{ \quad\\[1.0cm] Study of the {\Kpipi} decay }

\affiliation{Budker Institute of Nuclear Physics, Novosibirsk}
\affiliation{Chiba University, Chiba}
\affiliation{University of Cincinnati, Cincinnati, Ohio 45221}
\affiliation{Department of Physics, Fu Jen Catholic University, Taipei}
\affiliation{Justus-Liebig-Universit\"at Gie\ss{}en, Gie\ss{}en}
\affiliation{The Graduate University for Advanced Studies, Hayama}
\affiliation{Gyeongsang National University, Chinju}
\affiliation{Hanyang University, Seoul}
\affiliation{University of Hawaii, Honolulu, Hawaii 96822}
\affiliation{High Energy Accelerator Research Organization (KEK), Tsukuba}
\affiliation{Hiroshima Institute of Technology, Hiroshima}
\affiliation{University of Illinois at Urbana-Champaign, Urbana, Illinois 61801}
\affiliation{Institute of High Energy Physics, Chinese Academy of Sciences, Beijing}
\affiliation{Institute of High Energy Physics, Vienna}
\affiliation{Institute of High Energy Physics, Protvino}
\affiliation{Institute for Theoretical and Experimental Physics, Moscow}
\affiliation{J. Stefan Institute, Ljubljana}
\affiliation{Kanagawa University, Yokohama}
\affiliation{Korea University, Seoul}
\affiliation{Kyoto University, Kyoto}
\affiliation{Kyungpook National University, Taegu}
\affiliation{\'Ecole Polytechnique F\'ed\'erale de Lausanne (EPFL), Lausanne}
\affiliation{Faculty of Mathematics and Physics, University of Ljubljana, Ljubljana}
\affiliation{University of Maribor, Maribor}
\affiliation{University of Melbourne, School of Physics, Victoria 3010}
\affiliation{Nagoya University, Nagoya}
\affiliation{Nara Women's University, Nara}
\affiliation{National Central University, Chung-li}
\affiliation{National United University, Miao Li}
\affiliation{Department of Physics, National Taiwan University, Taipei}
\affiliation{H. Niewodniczanski Institute of Nuclear Physics, Krakow}
\affiliation{Nippon Dental University, Niigata}
\affiliation{Niigata University, Niigata}
\affiliation{University of Nova Gorica, Nova Gorica}
\affiliation{Osaka City University, Osaka}
\affiliation{Osaka University, Osaka}
\affiliation{Panjab University, Chandigarh}
\affiliation{Peking University, Beijing}
\affiliation{Princeton University, Princeton, New Jersey 08544}
\affiliation{RIKEN BNL Research Center, Upton, New York 11973}
\affiliation{Saga University, Saga}
\affiliation{University of Science and Technology of China, Hefei}
\affiliation{Seoul National University, Seoul}
\affiliation{Shinshu University, Nagano}
\affiliation{Sungkyunkwan University, Suwon}
\affiliation{University of Sydney, Sydney, New South Wales}
\affiliation{Tata Institute of Fundamental Research, Mumbai}
\affiliation{Toho University, Funabashi}
\affiliation{Tohoku Gakuin University, Tagajo}
\affiliation{Tohoku University, Sendai}
\affiliation{Department of Physics, University of Tokyo, Tokyo}
\affiliation{Tokyo Institute of Technology, Tokyo}
\affiliation{Tokyo Metropolitan University, Tokyo}
\affiliation{Tokyo University of Agriculture and Technology, Tokyo}
\affiliation{Toyama National College of Maritime Technology, Toyama}
\affiliation{Virginia Polytechnic Institute and State University, Blacksburg, Virginia 24061}
\affiliation{Yonsei University, Seoul}
  \author{I.~Adachi}\affiliation{High Energy Accelerator Research Organization (KEK), Tsukuba} % KEK
  \author{H.~Aihara}\affiliation{Department of Physics, University of Tokyo, Tokyo} % Tokyo
  \author{D.~Anipko}\affiliation{Budker Institute of Nuclear Physics, Novosibirsk} % BINP
  \author{K.~Arinstein}\affiliation{Budker Institute of Nuclear Physics, Novosibirsk} % BINP
  \author{T.~Aso}\affiliation{Toyama National College of Maritime Technology, Toyama} % Toyama
  \author{V.~Aulchenko}\affiliation{Budker Institute of Nuclear Physics, Novosibirsk} % BINP
  \author{T.~Aushev}\affiliation{\'Ecole Polytechnique F\'ed\'erale de Lausanne (EPFL), Lausanne}\affiliation{Institute for Theoretical and Experimental Physics, Moscow} % ITEP
  \author{T.~Aziz}\affiliation{Tata Institute of Fundamental Research, Mumbai} % Tata
  \author{S.~Bahinipati}\affiliation{University of Cincinnati, Cincinnati, Ohio 45221} % Cincinnati
  \author{A.~M.~Bakich}\affiliation{University of Sydney, Sydney, New South Wales} % Sydney
% ...
% BINP
\collaboration{The Belle Collaboration}

\begin{abstract}
We present a study of {\Kpipi} decay using 
$\sim$669 {\fbi} data, collected with the Belle detector at the KEKB asymmetric--energy $e^+ e^-$ collider.
The data is recorded at a center--of--mass energy 10.58 GeV.
The result for the branching ratio is  :
\begin{displaymath}
\mathcal{B} = (3.25 \pm 0.02(stat.) ^{+0.16} _{-0.15} (sys.)) \times 10^{-3}~.
\end{displaymath}
We also present results of the precise measurement of the branching ratio of other 3--prong decay modes, {\pipipi}, {\KKpi}, and {\KKK}.

\end{abstract}

\pacs{13.35.Dx, 12.15.Hh, 14.60.Fg}.

\maketitle

\tighten

{\renewcommand{\thefootnote}{\fnsymbol{footnote}}}
\setcounter{footnote}{0}

\section{Introduction}
\label{Section:Introduction}
\begin{figure}[htb]
\begin{center}
\includegraphics[width=0.6\textwidth]{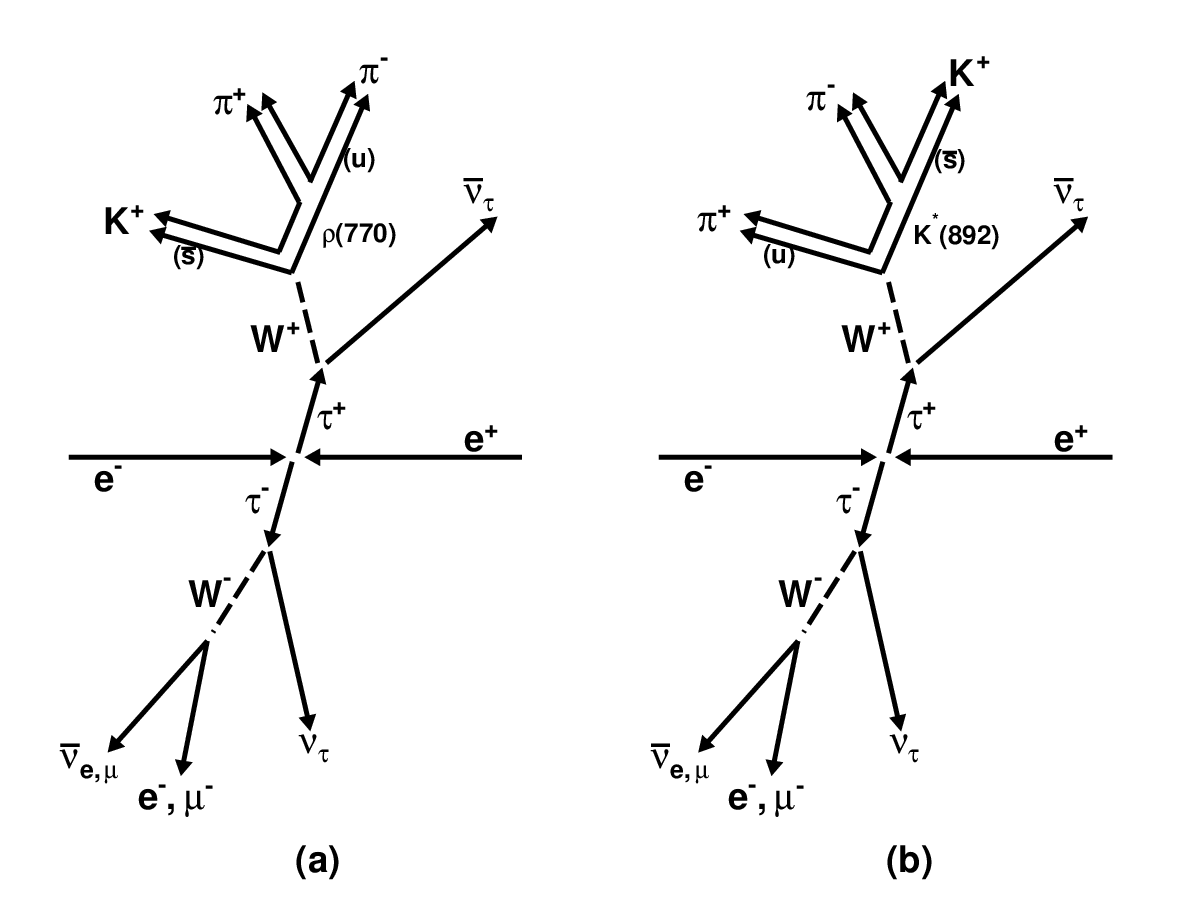}
\caption[Event shape]{A schematic view of the {\eetautau} in the center--of--mass system, where one $\tau$ ($\tau^+$) decays to the signal mode ($\tau^+\rightarrow K^+\pi^-\pi^+\nu_{\tau}$) and the other $\tau$ ($\tau^-$) decays to the pure leptonic modes ($\tau^-\rightarrow e^-\nu_{\tau}\overline{\nu}_{e}$ or $\tau^-\rightarrow \mu^-\nu_{\tau}\overline{\nu}_{\mu}$). Each figure shows one of two possible intermediate resonance states in the signal mode. }
\label{EventShape}
\end{center}
\end{figure}

The tau decay into 3 pseudoscalar particles has been studied since the discovery of the tau lepton. 
Moreover, decays to final states containing kaons provide a direct determination of the strange quark mass and Cabbibo--Kobayashi--Maskawa (CKM) matrix element $|V_{us}|$  
{\update{\cite{Citation:VusMs,Citation:ALEPH,Citation:OPAL}}}. 
Although the measurements of the {\Kpipi} decay from CLEO used sufficiently large statistics, 
the overall uncertainty is as large as $\sim$10\% \cite{Citation:CLEO}. 
{\updatea{Furthermore, the branching ratio of this decay, recently measured by BABAR \cite{Citation:BABAR}, is visibly smaller than the previous world average value \cite{Citation:PDG}.}}
The resonant states in {\Kpipi} decay have been studied only with poor statistics by CLEO \cite{Citation:CLEOResonance}. 
The resonance study is itself of interest for strange spectral function measurements and leptonic {\update{$CP$}} violation studies. 
Theoretically, two intermediate resonances, the $\rho(770)$ and $K^{*0}(892)$, are both expected to contribute to tau decay into $K^-\pi^+\pi^-\nu_{\tau}$:
$\tau^{-}\rightarrow K^{-} \rho^{0}(770) (\rightarrow \pi^{+} \pi^{-}) \nu_{\tau}$ and 
$\tau^{-}\rightarrow K^*(892)^0 (\rightarrow K^{-} \pi^{+}) \pi^{-} \nu_{\tau}$ \cite{Citation:Resonance}.
These are shown in \ref{EventShape}(a) and (b), respectively.
 
In this study, we present new measurements of the branching ratios for {\pipipi}, {\Kpipi}, {\KKpi}, and {\KKK} decays (Unless specified, the charge conjugation decay is also implied throughout this paper). The measurements of these kaon containing 3--prong decay modes and {\pipipi} decay are correlated due to particle misidentification, so they should be treated simultaneously.

This study uses a huge amount of tau decays collected with the Belle detector at the KEKB asymmetric--energy $e^+e^-$ collider \cite{Citation:KEKB}.
Since 1999, the Belle experiment has accumulated approximately 850 {\fbi} data until the summer of 2008. 
The Belle detector is a large-solid-angle magnetic
spectrometer that consists of a silicon vertex detector (SVD),
a 50-layer central drift chamber (CDC), an array of
aerogel threshold Cherenkov counters (ACC), 
a barrel-like arrangement of time-of-flight
scintillation counters (TOF), and an electromagnetic calorimeter {\update{(ECL)}}
comprised of CsI(Tl) crystals located inside
a super-conducting solenoid coil that provides a 1.5~T
magnetic field.  An iron flux-return located outside of
the coil is instrumented to detect $K_L^0$ mesons and to identify
muons (KLM).  The detector
is described in detail elsewhere~\cite{Citation:Belle}. 
In this study, we used  668.9 {\fbi} data collected on the $\Upsilon$(4S) resonance, 10.58 GeV, and 60 MeV below it (off--resonance), which corresponds to the production of $6.15\times10^8$ $\tau$--pairs.
 
\section{Selection of events}
\label{Section:SelectionOfEvent}

To select {\Kpipi} and other 3 prong decays, {\pipipi}, {\KKpi} and {\KKK}, first, tau pair events are selected. 
We require the number of tracks to be four 
and the sum of charges of these tracks to be zero, 
when the transverse momentum of a track in the laboratory frame is greater than 0.1 {\GeVc},  
and the track extrapolates to the interaction point within {\plusminus}1 cm transversely and {\plusminus}5 cm along the beam direction. 
The sum of the reconstructed momenta in the center--of--mass (CM) frame is required to be less than 10 {\GeVc}, 
and the sum of energy deposited in the calorimeter is required to be less than 10 GeV.
The maximum transverse momentum is required to be greater than 0.5 {\GeVc}, to reject two--photon events which include many low transverse momentum tracks. 
To reject beam--related background, we cut on the position of the reconstructed event vertex, 
requiring it to be closer to the interaction point {\plusminus}0.5 cm transversely and {\plusminus}3 cm along the beam direction. 
The missing mass $M_{{\update{\rm {miss}}}}^2 = ( {\it p}_{\rm {init}} - \sum_{\rm {tr}}{\it p_{\rm {tr}}} - \sum_{\gamma}{\it p_{\gamma}} )^2$ 
and {\update{the polar angle of missing momentum}} in the CM frame are efficient variables for rejecting two--photon and Bhabha backgrounds.
In the definition of missing mass, ${\it p_{\rm {tr}}}$ and ${\it p_{\gamma}}$ are the four--momenta of measured tracks and photons, respectively,
and ${\it p}_{\rm {init}}$ is the initial CM frame momentum of the $e^+e^-$ beams. 
We require that the missing mass should be greater than 1 {\GeVcc} and less than 7 {\GeVcc}, and the polar angle with respect to the beam direction 
in the CM frame should be greater than 30{\degree} and less than 150{\degree}. 

\begin{figure}[htb]
\begin{center}
\includegraphics[width=0.6\textwidth]{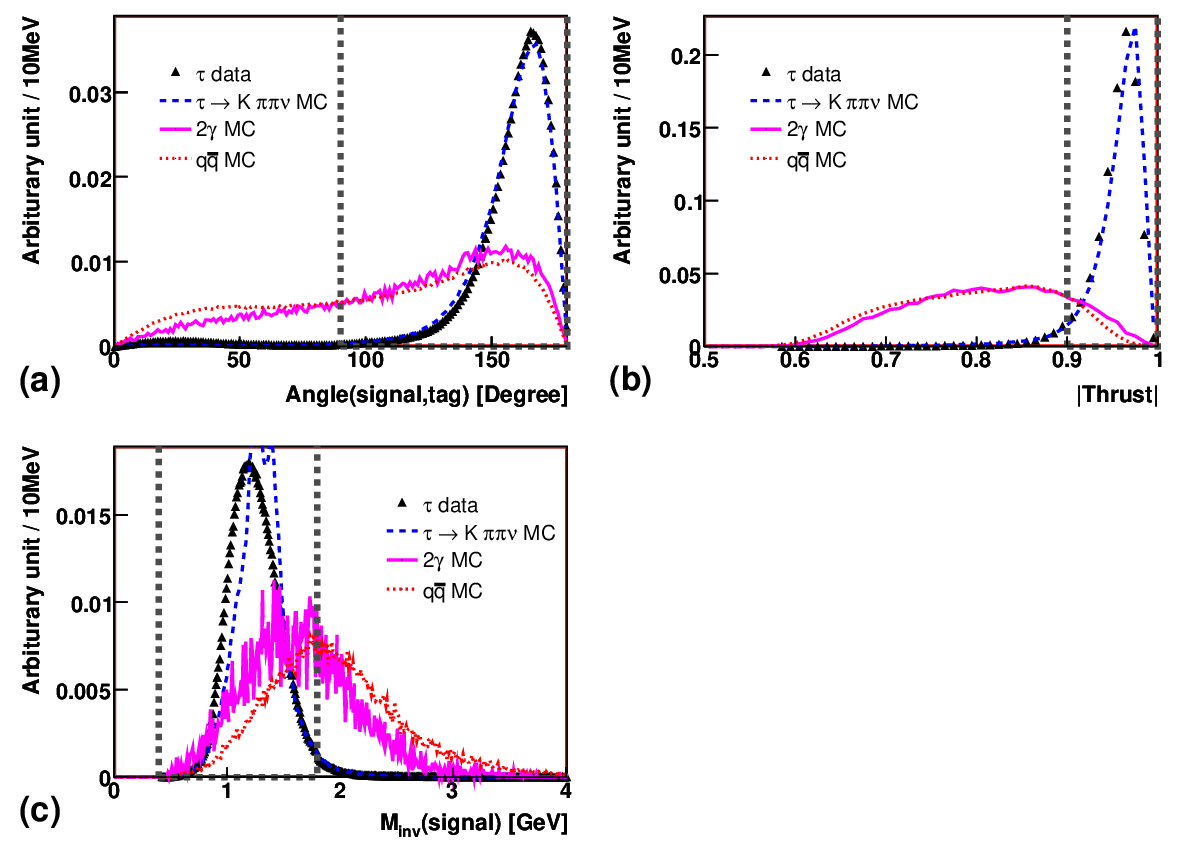}
\caption[Performance of cut]{Performance of event selection criteria.  
{\update{(a) The angle between the total momenta of the hadronic system on the signal side and the lepton momentum on the tag side in the CM system.
(b) The magnitude of the thrust.
(c) The invariant mass of the hadronic system.}}
The triangular points show data, the dashed lines show {\Kpipi} signal MC, solid lines show two--photon background, and dotted lines show {\eeqq} background. 
{\update{The number of events of each sample is normalized to be the same.}}}
\label{CutStatus}
\end{center}
\end{figure}

Particle identification 
is performed to select tau events containing one lepton (electron or muon) and 3 hadrons (pions or kaons). 
The magnitude of thrust is evaluated and is required to be greater than 0.9, to suppress two--photon and {\eeqq} backgrounds, 
where the thrust is defined by the maximum of $(\sum_i | \hat{n} \cdot \vec{p_i}|) / (\sum_i |\vec{p_i}|)$ 
when $\vec{p_i}$ is the momentum of $i$--th track and 
where $\hat{n}$ is the unit vector in the direction of the thrust axis -- the direction maximizing the thrust. 
{\update{We require that the angle between the total momenta of the hadrons and the lepton momentum in the CM system should be greater than 90{\degree},  
{\updatea{whereby}} the tag side lepton and signal side hadrons lie in opposite hemispheres, 
the so--called 1--3 prong configuration.}}
Also the invariant mass of charged tracks and gamma clusters for each side are required to be less than the tau mass. 
Finally we require that there are no {\Ks}, {\pizero}, and energetic $\gamma$ on the signal side.
{\update{Figure}} \ref{CutStatus} shows the performance of the criteria used to select the 1--3 prong sample, 
{\update{where the required conditions are shown by the vertical lines in each figure.}}
The events finally selected are candidates for $\tau^-\rightarrow h_1^-h_2^+h_3^-\nu_{\tau}$,
where $h_{1,2,3}$ is a hadron identified as either pion or kaon. 
At this stage, the reconstruction efficiency of {\Kpipi} decay is $\sim28\%$, 
while the dominant background modes, {\pipipipizero}, {\Kspi}, and {\eeqq} are 
reduced to 6.0\%, 1.8\%, and 0.004\%, respectively. 
{\update{The reconstruction efficiency of two--photon background is {\updatea{so small ($1.6 \times 10^{-4}$ \%) that it is}} negligible.}}
 
\begin{figure}[htb]
\begin{center}
\includegraphics[width=0.6\textwidth]{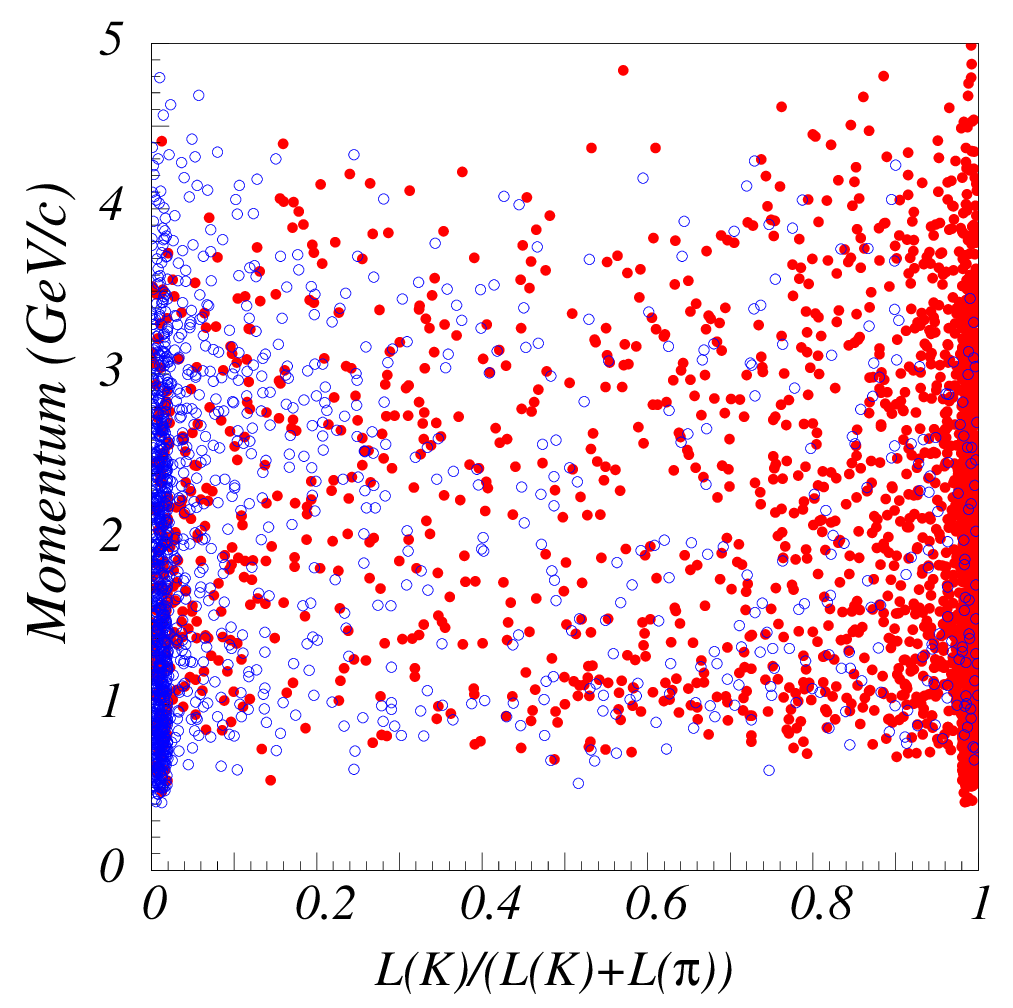}
\caption[Kaon ID likelihood ratio] {Likelihood ratio of kaon identification, $L(K)/(L(K)+L(\pi))$, versus the momentum of the particles. The red filled circles and blue empty circles represent the real kaons and pions, respectively.} 
\label{KidLikelihoodRatio}
\end{center}
\end{figure}

One of the most important issues in this analysis is the separation of kaons and pions. 
In the Belle experiment, $dE/dx$ information from the CDC, hit information of the ACC, and time--of--flight from the TOF are used to 
construct the {\update{likelihood for kaon (pion) hypothesis, $L(K)$ ($L(\pi)$)}}. 
{\update{Figure}} \ref{KidLikelihoodRatio} shows the likelihood ratio of kaon identification, $L(K)/(L(K)+L(\pi))$, for the charged particles versus their momenta.
A clean separation between kaon and pion is observed in the whole momentum region relevant for this analysis, 
{\update{where the momentum of kaon or pion ranges from 0.1 {\GeVc} to $\sim$5 {\GeVc}, and the average of momentum is $\sim$1.3 {\GeVc}}}.
As is discussed later, we choose a relatively tight particle identification (PID) condition for the kaon ($L(K)/(L(K)+L(\pi)) > 0.9$) and a relatively loose condition for the pion {\updatea{($L(K)/(L(K)+L(\pi)) < 0.9$)}}.  
With these criteria, the efficiency of kaon identification is $\sim 73\%$, and the fake rate of pion mis--identification to kaon is $\sim 5\%$.
The calibration of the kaon and pion identification efficiencies and their fake rates are carried out
by the data using the kaon and pion tracks in the $D^{*+}$ decays,
$D^{*+} \rightarrow D^0 (\rightarrow K^- \pi^+) \pi^+_s$ sample, 
where one knows which tracks are kaons and pions from kinematics and the charges of the tracks.
We evaluate the efficiencies and fake rates for this calibration sample and compare them to the Monte Carlo (MC) expectations.
From this comparison, we prepare a correction table as a function of track momenta and polar angles, and apply them to the MC.

\begin{figure}[htb]
\begin{center}
\includegraphics[width=0.6\textwidth]{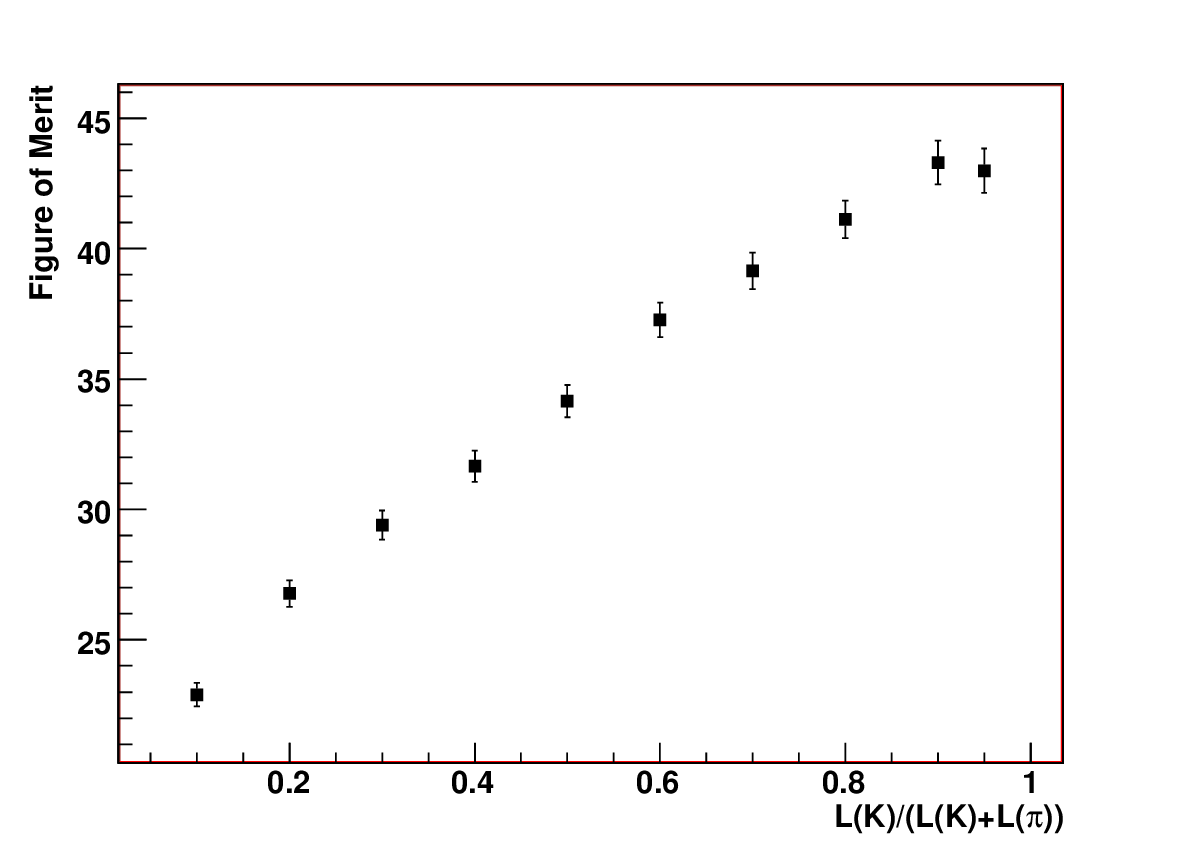}
\caption[PID study result]{ {\updatea{Figure-of-merit as a function of the kaon likelihood ratio value $L(K)/(L(K)+L(\pi))$ {\updateb{in}} which a particle is assigned as a kaon and otherwise assigned as a pion}}. {\update{The uncertainties are evaluated from the comparison of the efficiencies and fake rates of kaon identification with that of control sample, $D^{*+} \rightarrow D^0 (\rightarrow K^- \pi^+) \pi^+_s$~.}} }
\label{PIDstudy1}
\end{center}
\end{figure}

This kaon identification criterion is determined by maximizing the figure--of--merit (FOM), 
{\update{when the requirements of kaon identification likelihood ratio is varied}}.
The figure--of--merit is defined as:
\begin{equation}
FOM = \frac{S}{\sqrt{S+N}}~,
\end{equation}
{\updatea{where $S$ is the number of the signal ({\Kpipi}), 
and $N$ is the number of cross--feed background ({\pipipi} and {\KKpi})}}.
The result of FOM is shown in Fig. \ref{PIDstudy1}, where one can see that the FOM is maximal with the particle identification criteria used in this analysis.  
For the case of electron and muon identification, the probabilities to be an electron or muon are evaluated using the information from ECL, KLM, and other particle identification detectors. Also the efficiencies and the systematic uncertainties of lepton identification are evaluated by using the control sample $\gamma\gamma\rightarrow e^+e^-/\mu^+\mu^-$. 

To reduce the feed-down background which contain {\Ks} and {\pizero}, 
we reconstruct {\Ks} and {\pizero} signals explicitly.
{\Ks} is reconstructed from two charged pion tracks having invariant mass $M(\pi\pi)$ within {\plusminus}13.5 {\MeVcc} of the {\Ks} mass.
To improve the purity of {\Ks},
the point of closest approach to the interaction point along the extrapolation of each track
is required to be greater than 0.3 cm transverse to the beam direction. 
The azimuthal angle between the momentum vector and the decay vertex vector of reconstructed {\Ks} is required to be less than 0.1 rad. 
The distance between the two daughter pion tracks at their interception point is required to be less than 1.8 cm, and the flight length of {\Ks} transversely to the beam direction is required to be greater than 0.08 cm. 
Finally we required that the invariant mass of a {\Ks} candidate should not be located in the $\Lambda$, $\bar{\Lambda}$, and $\gamma$ mass range, 
with the daughter tracks assumed to be electrons or pions and protons or anti--protons as appropriate.
Also the {\pizero}s are reconstructed from two gamma clusters. The energy of each gamma is required to exceed 50 MeV for the candidates in the barrel part of the calorimeter, and 100 MeV for the candidates in the endcap part of the calorimeter. Also we applied a requirement on the invariant mass of two gammas $M(\gamma\gamma)$ to be within {\plusminus}16.2 {\MeVcc} of the {\pizero} mass. 
An additional selection that there should be no energetic gamma whose energy is greater than 0.3 GeV, is applied to reject remaining events containing {\pizero}. 

\section{Efficiency estimation}
\begin{figure}[htb]
\begin{center}
\includegraphics [width=0.6\textwidth]{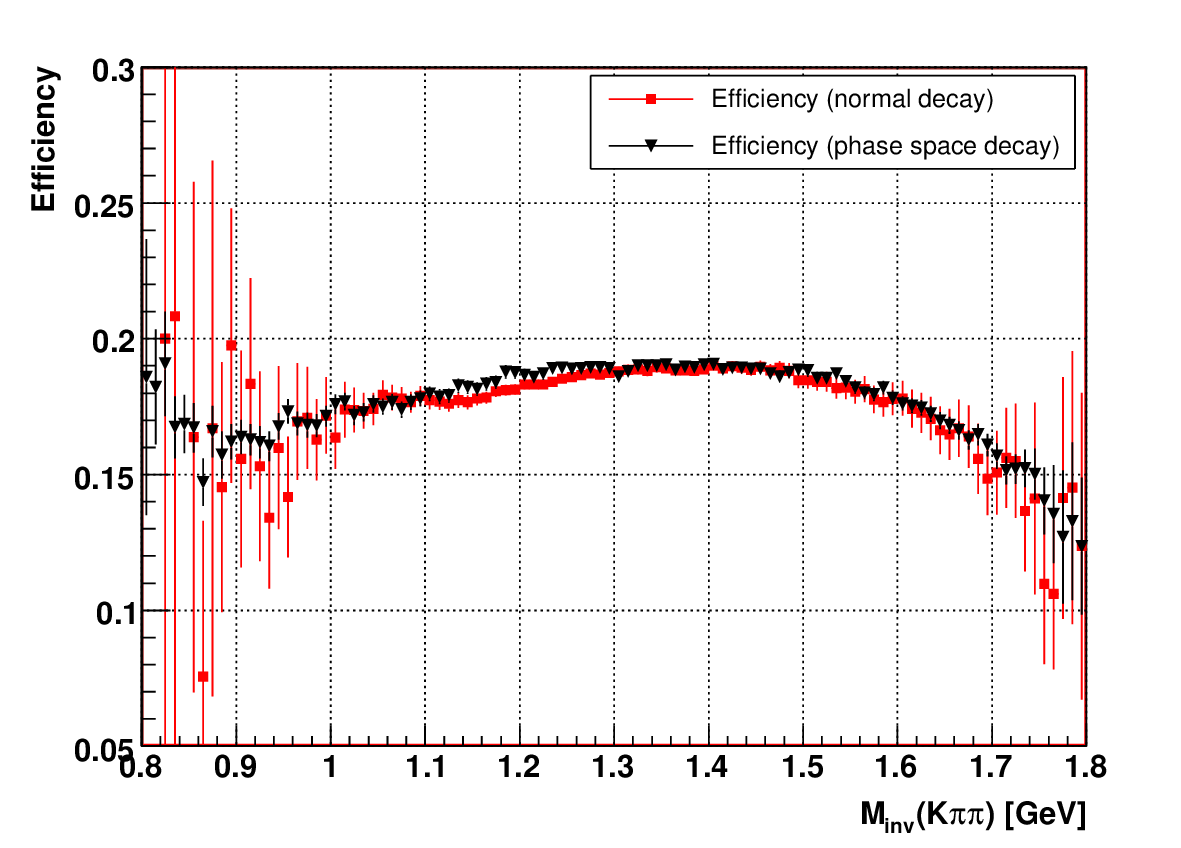}
\caption[Efficiency of different decay model]{The efficiency of the {\Kpipi} decay as a function of {\MKpipi}. Squares and triangular points represent the normal decay {\updatea{model}}, and the phase space decay {\updatea{model}}, respectively.}
\label{EfficiencyCurve}
\end{center}
\end{figure}

We generated $\sim$5 {\update{million}} signal mode MC data with the TAUOLA \cite{Citation:TAUOLA} based KKMC \cite{Citation:KKMC} generator, 
to estimate signal efficiency.
Also the detector response of all MC data sets is simulated with the GEANT \cite{Citation:GEANT} simulator. 
The TAUOLA generator uses a decay model for {\Kpipi}, where the process is mediated by $K_1(1270)$ and $K_1(1410)$ resonances and their interference 
(we will call this decay model as ``normal'' decay model, in the following).
To check for a possible bias of efficiency due to the specific decay model used, 
events assuming phase--space decay of {\Kpipi} were also generated using the KKMC program.
The evaluated efficiencies for {\MKpipi} for both cases are compared in Fig. \ref{EfficiencyCurve}. 
The relative difference of {\updatea{efficiencies}} between {\updateb{the normal decay and the phase space decay}} is around 1 \%, so small that it is negligible. 

The average efficiencies and the {\updatea{fractions}} of the {\update{cross--feed}} background for all 3--prong decays are summarized in Table \ref{EfficiencyTable}, 
where we used the normal decay model for the determination of the efficiency. 
One can see that the fake rate of cross--feed from {\pipipi} to {\Kpipi}, 
is very small, but since the branching ratio of {\pipipi} is large, there is a substantial contamination in the reconstruction of {\Kpipi} events coming from misidentified {\pipipi} events. 

\begin{center}
\begin{table}[!hbp]
\caption{\update{Summary of the efficiencies and the {\updatea{fractions}} of cross--feed.}}
\begin{tabular}{l|c|c|c|c}
\hline
 & \multicolumn{4}{c}{Generated decay mode} \\
\cline{2-5} 
Reconstructed 	& $\tau^-\rightarrow$ 		&$\tau^-\rightarrow$  		& $\tau^-\rightarrow$ 		&  $\tau^-\rightarrow$ \\
decay mode 	& $\pi^-\pi^+\pi^-\nu_{\tau}$	& $K^-\pi^+\pi^-\nu_{\tau}$	&$K^-K^+\pi^-\nu_{\tau}$	& $K^-K^+K^-\nu_{\tau}$ \\
\hline
{\pipipi} & 0.23		 	& 0.076			& 0.023			& $7.3\times 10^{-3}$ 	\\
{\Kpipi}  & 0.012			& 0.17			& 0.049			& 0.023			\\
{\KKpi}   & $4.0\times 10^{-4}$	& $4.7\times 10^{-3}$	& 0.13			& 0.060			\\
{\KKK}    & $2.8\times 10^{-6}$	& $1.4\times 10^{-4}$	& $2.8\times 10^{-3}$	& 0.094			\\
\hline
\end{tabular}
\label{EfficiencyTable}
\end{table}
\end{center}

\section{Branching ratio calculation}
\label{Section:CalculationOfBranchingRatio}

{\update{After applying all selection criteria, the number of the events {\updatea{surviving}} as 
{\Kpipi}, {\pipipi}, {\KKpi}, and {\KKK} are summarized in Table \ref{NumberOfEvents1}. 
The possible {\updatea{background contaminations}} in these signals are from (a) cross-feed from the signal modes and
(b) other processes such as {\pipipipizero}, {\Kspi}, and {\eeqq}.
The fraction of the background coming from the other processes is 5\% to 10\%, 
as summarized in the Table \ref{NumberOfEvents1} (3rd column) for each signal mode.
The 4th column shows the fraction of the the main  background among them:
The decay with {\pizero}, {\pipipipizero} is dominant one for the {\pipipi} and {\Kpipi} modes,
while the background from {\updatea{the}} continuum process {\eeqq} is {\updatea{dominant}} for the {\KKpi} and {\KKK} modes.}}

To take into account the cross-feeds between the decay channels, the true number of the yield 
$N_{i}^{\rm {true}}$ ($i =$ {\pipipi}, {\Kpipi}, {\KKpi}, and {\KKK}), 
is obtained from the following equation
\begin{equation}
N^{\rm {true}}_i = \sum_j \mathcal{E}^{-1}_{ij} (N^{\rm {rec}}_j - N^{\rm {other}}_j)~,
\label{breq1}
\end{equation}
where $N_{j}^{\rm {rec}}$ is the number of the reconstructed $j$--th decay mode 
and $N_{j}^{\rm {other}}$ is the number of the background to the $j$--th mode coming from the other sources such as {\pipipipizero}~. 
$\mathcal{E}_{ij}$ is the efficiency for detecting mode $i$ as mode $j$ and $\mathcal{E}_{ij}^{-1}$ is the
inverse of the $\mathcal{E}_{ij}$ matrix.
The value of the efficiency (and migration) matrix $\mathcal{E}_{ij}$ determined by MC is given in Table \ref{EfficiencyTable}.

In order to determine the branching fraction, we use events of pure leptonic decays, where
one tau decays to {\enunu} and the other decays to {\mununu} {\update{\cite{Citation:Epifanov}}} (Hereafter we call such events {\emuevent} events).
The branching fraction for the decay mode $i$ can then be written as:
\begin{equation}
\mathcal{B}_i 
  = N^{\rm {true}}_{i} \cdot \frac{ \varepsilon_{e\mu} }{N_{{\update{\rm {sig}}},e\mu}  } \cdot \frac{\mathcal{B}_{\tau\rightarrow e\overline{\nu}\nu}\cdot \mathcal{B}_{\tau\rightarrow\mu\overline{\nu}\nu}}{\mathcal{B}_{\tau\rightarrow l\overline{\nu}\nu}}~, 
\end{equation}
where  $N_{ {\update{\rm{sig}}},e\mu}$ and $\varepsilon_{e\mu}$ are the number of 
{\emuevent} events and the corresponding detection efficiency, respectively. 
$\mathcal{B}_{\tau\rightarrow e\overline{\nu}\nu}$ and $\mathcal{B}_{\tau\rightarrow\mu\overline{\nu}\nu}$ are the branching fractions of {\enunu} and {\mununu} decays, respectively, 
and $\mathcal{B}_{\tau\rightarrow l\overline{\nu}\nu} = \mathcal{B}_{\tau\rightarrow e\overline{\nu}\nu} + \mathcal{B}_{\tau\rightarrow\mu\overline{\nu}\nu}$.

\begin{figure}[htb]
\begin{center}
\includegraphics[width=0.6\textwidth]{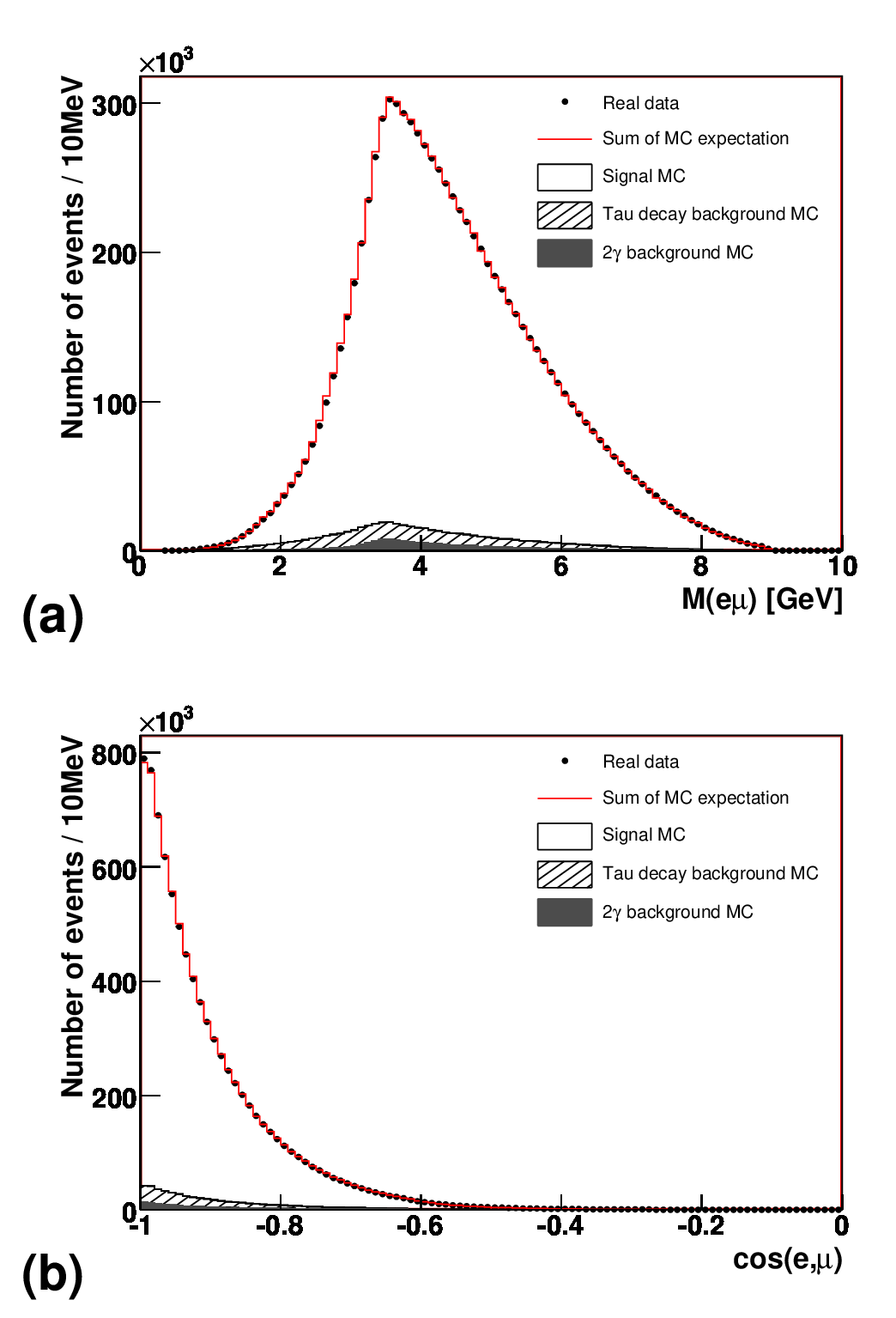}
\caption[{\emuevent} event selection (M(e$\mu$), \cos(e,$\mu$))]{{\updatea{Distributions}} of the {\emuevent} events{\updatea{:}} (a) Invariant mass of electron and muon system. (b) Cosine of the angle between electron and muon. The closed circles and the solid histogram represent the data and the sum of the MC expectation, respectively. The open area represent the {\emuevent} signal events, while the tau--pairs other than the {\emuevent} sample and the two--photon background are shown by the hatched and the gray histogram, respectively. }
\label{EmuSelection1}
\end{center}
\end{figure}

With this normalization method, it is important to measure the number of {\emuevent} events and the corresponding efficiency precisely. 
{\update{Figure}} \ref{EmuSelection1} shows the comparison of the invariant mass of electron and muon system and the cosine of the angle between electron and muon, for the real {\emuevent} events and the sum of MC expectation. The MC reproduces the data reasonably. 
By subtracting the backgrounds estimated by MC from the data, we {\update{measure}} the number of {\emuevent} events. 

\begin{center}
\begin{table}[htp]
\caption{Number of reconstructed events (second column), the fraction of the backgrounds other than 3--prong cross--feeds (third column), and the main source of other backgrounds with its fraction to the total other backgrounds (fourth column).}
\begin{tabular}{l|c|c|l}
\hline
 & $N^{\rm {rec}}$ & $N^{\rm {other}}/N^{\rm {rec}}$ (\%) & Main component in $N^{\rm {other}}$ \\
\hline
{\pipipi} & $7.856\times10^6$	& 10.61\%	& 64.3\% of {\pipipipizero}	\\
{\Kpipi} 	& $7.944\times10^5$	& 12.15\%	& 34.4\% of {\pipipipizero}	\\
{\KKpi} 	& $1.076\times10^5$	& 6.70\%	& 30.4\% of {\eeqq}	\\
{\KKK} 	& $3.162\times10^3$	& 5.45\%	& 53.1\% of {\eeqq}	\\
\hline
\end{tabular}
\label{NumberOfEvents1}
\end{table}
\end{center}

\begin{figure}[htb]
\begin{center}
\includegraphics[width=0.6\textwidth]{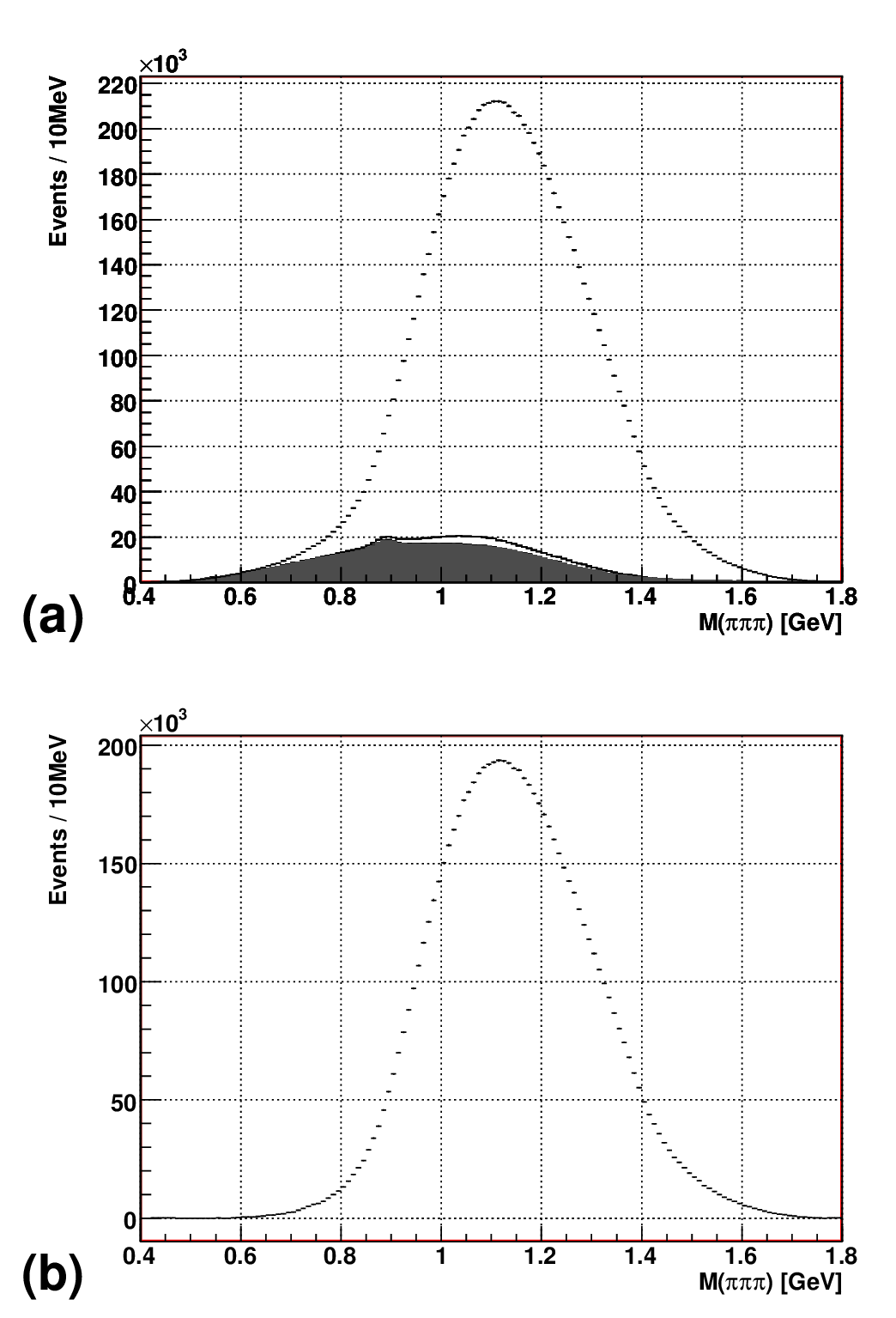}
\caption[{\Mpipipi} distribution of {\pipipi} mode]{(a) {\Mpipipi} distribution for {\pipipi} for data (black points). The open histogram is the cross--feed from {\Kpipi}, and the dark gray histogram is the sum of {\update{all other backgrounds}}. (b) {\Mpipipi} distribution with background subtraction.}
\label{pipipiResult}
\end{center}
\end{figure}

\begin{figure}[htb]
\begin{center}
\includegraphics[width=0.6\textwidth]{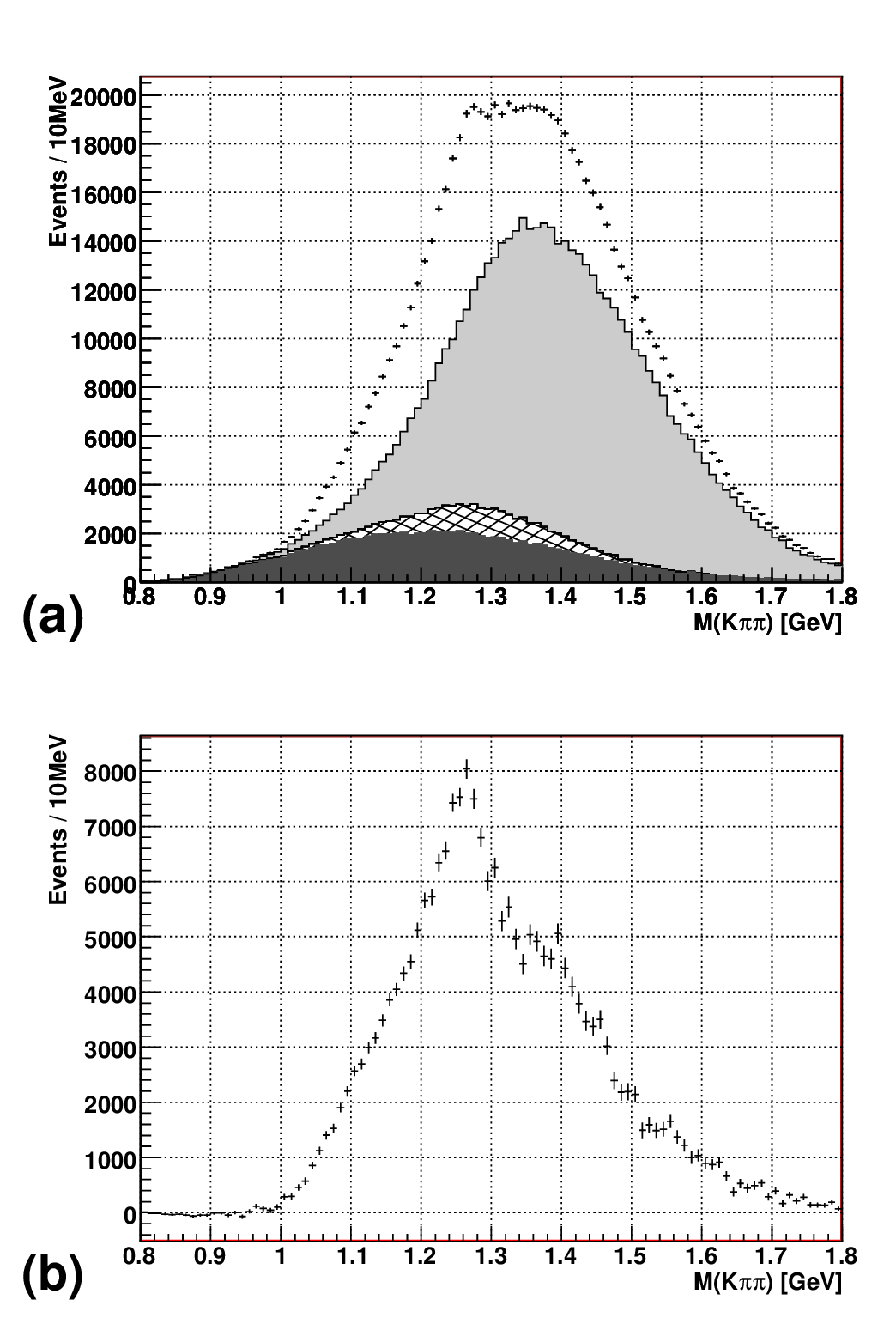}
\caption[{\MKpipi} distribution of {\Kpipi} mode]{(a) {\MKpipi} distribution for {\Kpipi} for data (black points). The light gray histogram is the cross--feed from {\pipipi}, the hatched histogram is the cross--feed from {\KKpi}, and the dark gray histogram is the sum of {\update{all other backgrounds}}. (b) {\MKpipi} distribution with background subtraction.}
\label{KpipiResult}
\end{center}
\end{figure}

\begin{figure}[htb]
\begin{center}
\includegraphics[width=0.6\textwidth]{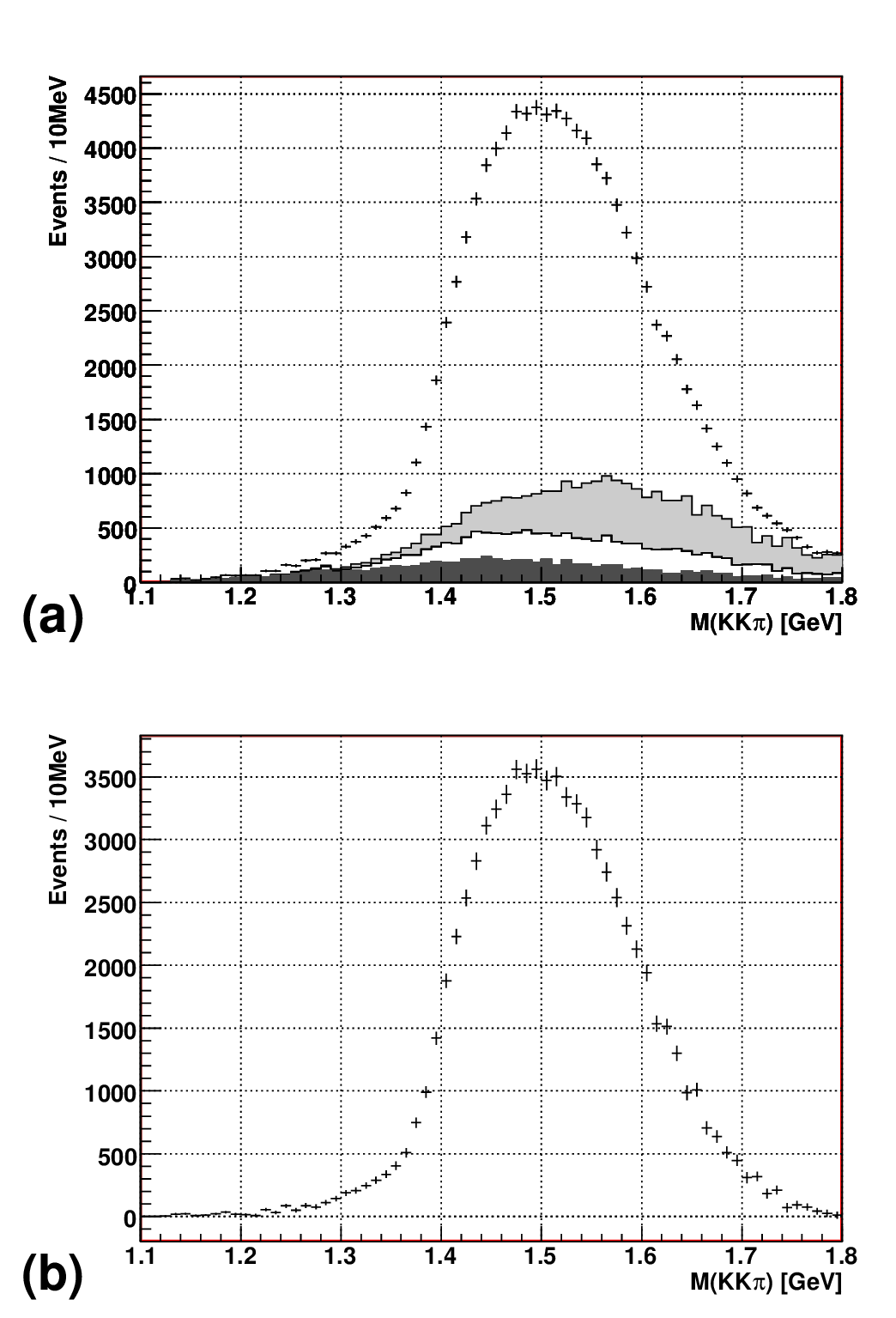}
\caption[{\MKKpi} distribution of {\KKpi} mode]{(a) {\MKKpi} distribution for {\KKpi} for data (black points). The light gray histogram is the cross--feed from {\pipipi}, the white open histogram is the cross--feed from {\Kpipi}, and the dark gray histogram is the sum of {\update{all other backgrounds}}. (b) {\MKKpi} distribution with background subtraction.}
\label{KKpiResult}
\end{center}
\end{figure}

\begin{figure}[htb]
\begin{center}
\includegraphics[width=0.6\textwidth]{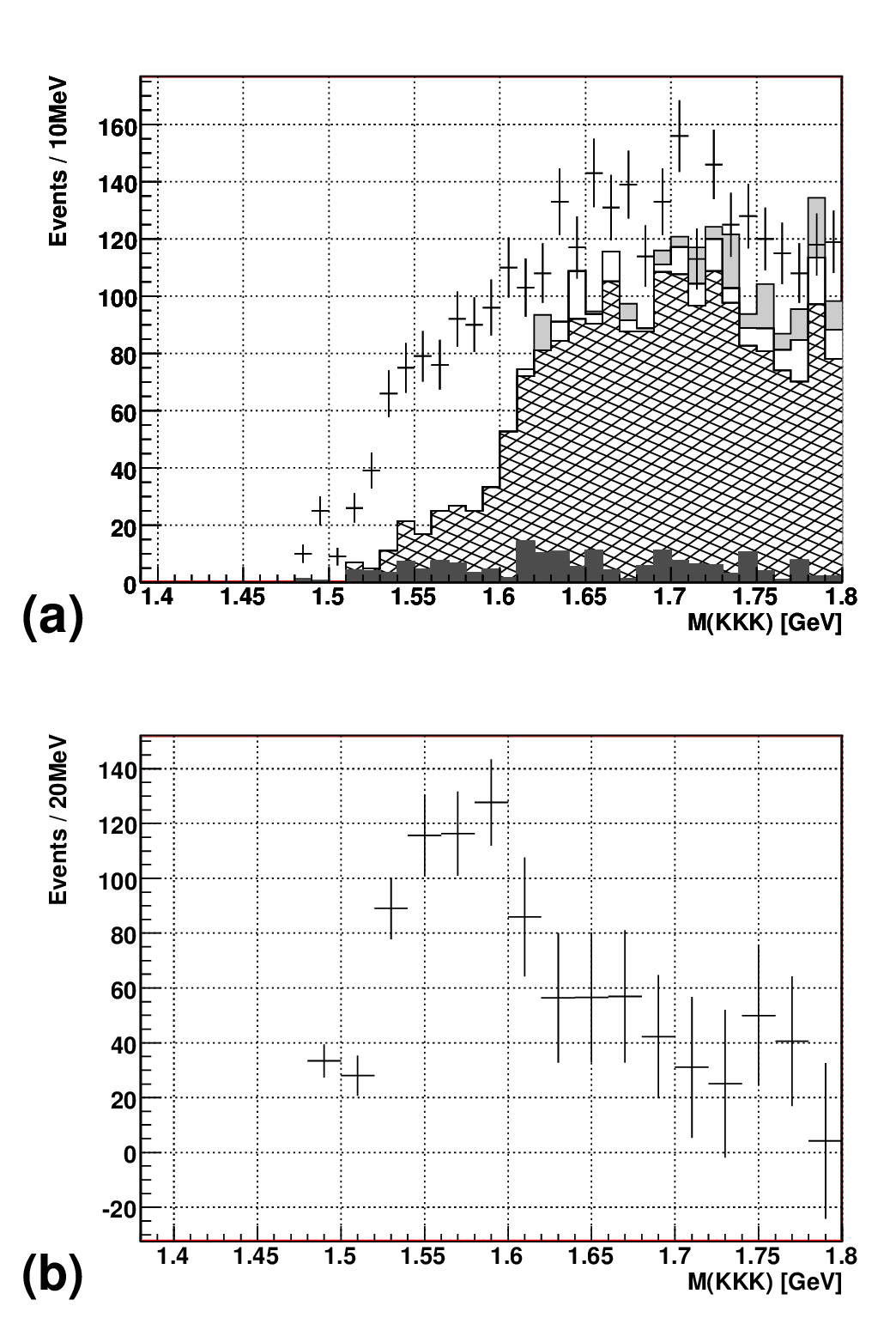}
\caption[{\MKKK} distribution of {\KKK} mode]{(a) {\MKKK} distribution for {\KKK} for data (black points). The light gray histogram is the cross--feed from {\pipipi}, the white open histogram is the cross--feed from {\Kpipi}, the hatched histogram is the cross--feed from {\KKpi}, and the dark gray histogram is the sum of {\update{all other backgrounds}}. (b) {\MKKK} distribution with background subtraction.}
\label{KKKResult}
\end{center}
\end{figure}

{\updatea{Figures}} \ref{pipipiResult}, \ref{KpipiResult}, \ref{KKpiResult}, and \ref{KKKResult} show the comparison of the invariant mass distribution {\Mpipipi}, {\MKpipi}, {\MKKpi}, and {\MKKK} for the data and the backgrounds estimated from the MC. The cross--feed background components are re--scaled using the branching ratio value calculated above.  

Table \ref{Systematics} summarizes the various contributions to the systematic uncertainties on the branching ratio.
For all cases, the uncertainty of the track finding efficiency is the most important source of systematic uncertainty, 
which is estimated from the comparison of real and MC data for $D^* \rightarrow \pi D^0$ and $D^0 \rightarrow \pi\pi K_S^0(K_S^0 \rightarrow \pi^+ \pi^-)$ decay sample. 
The uncertainty of the efficiency migration matrix includes that of the kaon identification efficiency and lepton identification efficiency, which are important sources of the systematic errors.
The uncertainty of the kaon identification efficiency is evaluated from the comparison of real and MC data for $D^{*+} \rightarrow D^0 \pi^+_s$, and $D^0 \rightarrow K^- \pi^+$ events, while 
$\gamma \gamma \rightarrow e^+ e^- / \mu^+ \mu^-$ events are used for the estimation of lepton identification efficiency uncertainty. 
The effect of the uncertainty of luminosity and cross section of {\eetautau} {\update{\cite{Citation:TauTauCrosssection}}}  are rather small, because we used {\emuevent} events for the normalization. 
Trigger efficiency for 3--prong modes is $\sim$ 86 \%, and its fluctuation is estimated to be $\sim$0.6 \%, using the trigger simulation program for the signal decays. 
The uncertainty due to gamma veto is evaluated by using the different selection criteria of gamma energy. 
The uncertainty on the background subtraction is evaluated from the error propagation of the branching ratio of tau decay modes other than 3--prong modes. 
Also the uncertainty of branching ratio of the leptonic decay of tau is taken into account. 

\begin{center}
\begin{table}[htp]
\caption{Summary on the systematical uncertainties.}
\begin{tabular}{l|c|c|c|c}
\hline
 & $\tau \rightarrow$	& $\tau \rightarrow$	& $\tau \rightarrow$	& $\tau \rightarrow$ 	\\
 & $\pi\pi\pi\nu$	& $K\pi\pi\nu$		& $KK\pi\nu$		& $KKK\nu$		\\\hline
Tracking efficiency error & +3.2/-3.0 & +3.2/-3.0 & +3.2/-3.0 & +3.1/-2.9 \\
Efficiency matrix and PID & +1.5/-1.5 & +1.7/-1.7 & +1.9/-1.9 & +2.3/-2.3 \\
Trigger efficiency & +0.5/-0.5 & +0.5/-0.5 & +0.6/-0.6 & +0.6/-0.6 \\
Luminosity & +0.1/-0.1 & +0.1/-0.1 & +0.1/-0.1 & +0.1/-0.1 \\
Gamma veto & +0.8/-0.8 & +2.5/-2.5 & +1.0/-1.0 & +0.9/-0.9 \\
Background estimation & +0.3/-0.3 & +2.0/-2.0 & +0.2/-0.2 & +0.3/-0.3 \\
Branching ratio of leptonic decay & +0.2/-0.2 & +0.2/-0.2 & +0.2/-0.2 & +0.2/-0.2 \\\hline
Total (\%) & +3.7/-3.6 & +4.9/-4.7 & +3.9/-3.7 & +4.0/-3.9 \\\hline
\end{tabular}
\label{Systematics}
\end{table}
\end{center}

After taking into account the backgrounds, the efficiencies and the various sources of systematic errors discussed above, we obtain the branching ratio for {\Kpipi} decay, 
\begin{displaymath}
\mathcal{B} = (3.25 \pm 0.02(stat.) ^{+0.16} _{-0.15} (sys.)) \times 10^{-3}~.
\end{displaymath}
This value agrees well with the current world average.
The branching ratios for other 3--prong decays are summarized in Table \ref{BranchingRatio}. 
For all modes, {\updateb{the accuracy is improved significantly especially for modes including kaons}},
and the branching ratios are consistent with {\updatea{the}} {\update{previous}} world average {\updatea{values}} \cite{Citation:PDG}.
But there are some discrepancies with the most recent precise measurement from BABAR, 
{\update{where the result of branching ratio of {\Kpipi} decay from BABAR is $\mathcal{B} = (2.73 \pm 0.02(stat.) \pm 0.09(sys.)) \times 10^{-3}$ \cite{Citation:BABAR}}}.

\begin{center}
\begin{table}[htp]
\caption{Summary of the branching ratios.}
\begin{tabular}{l|c}
\hline
 & Branching ratio \\\hline
{\pipipi} & $(8.41 \pm 0.00(stat.) ^{+0.34} _{-0.33} (sys.)) \times 10^{-2}$  \\
{\Kpipi} & $(3.25 \pm 0.02(stat.) ^{+0.16} _{-0.15} (sys.)) \times 10^{-3}$  \\
{\KKpi} & $(1.53 \pm 0.01(stat.) ^{+0.06} _{-0.06} (sys.)) \times 10^{-3}$  \\
{\KKK} & $(2.60 \pm 0.23(stat.) ^{+0.10} _{-0.10} (sys.)) \times 10^{-5}$  \\\hline
\end{tabular}
\label{BranchingRatio}
\end{table}
\end{center}

\section{Conclusion}

\begin{figure}[htb]
\begin{center}
\includegraphics[width=0.6\textwidth]{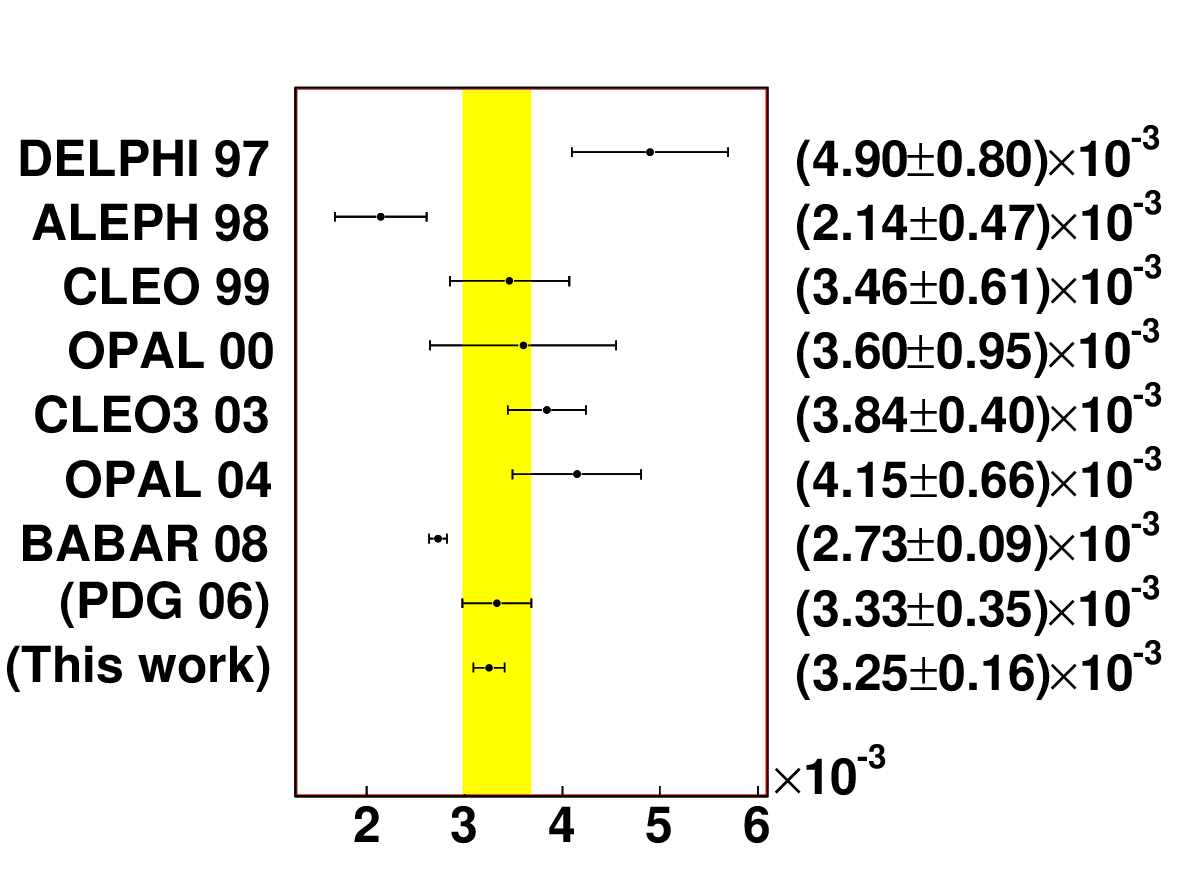}
\caption[{\update{Summary of {\Kpipi} decay branching ratio measurements}}]{{\update{Summary of {\Kpipi} branching ratio measurements.}}}
\label{KpipiBranchingRatioHistory}
\end{center}
\end{figure}

We present results from the measurement of the branching ratio for the {\Kpipi} decay mode, 
as well as those for the {\pipipi}, {\KKpi}, and {\KKK} decay modes. 
{\update{The results are based on the 669 {\fbi} $\tau$ pair sample collected by the Belle detector.}}
The result for the {\Kpipi} decay mode is compared with previous measurements in Fig. \ref{KpipiBranchingRatioHistory}. 
The statistical and systematic uncertainties are compatible with a recent precision measurement from BABAR, 
although the branching ratio of this study is more consistent with {\updatea{the}} {\update{previous}} world average {\updatea{values}}. 

\section{Acknowledgments}
We thank the KEKB group for the excellent operation of the
accelerator, the KEK cryogenics group for the efficient
operation of the solenoid, and the KEK computer group and
the National Institute of Informatics for valuable computing
and SINET3 network support. We acknowledge support from
the Ministry of Education, Culture, Sports, Science, and
Technology of Japan and the Japan Society for the Promotion
of Science; the Australian Research Council and the
Australian Department of Education, Science and Training;
the National Natural Science Foundation of China under
contract No.~10575109 and 10775142; the Department of
Science and Technology of India; 
the BK21 program of the Ministry of Education of Korea, 
the CHEP SRC program and Basic Research program 
(grant No.~R01-2005-000-10089-0) of the Korea Science and
Engineering Foundation, and the Pure Basic Research Group 
program of the Korea Research Foundation; 
the Polish State Committee for Scientific Research; 
the Ministry of Education and Science of the Russian
Federation and the Russian Federal Agency for Atomic Energy;
the Slovenian Research Agency;  the Swiss
National Science Foundation; the National Science Council
and the Ministry of Education of Taiwan; and the U.S.\
Department of Energy.

\end{document}